\documentclass[conference]{IEEEtran}
\IEEEoverridecommandlockouts
\usepackage{cite}
\usepackage{amsmath,amssymb,amsfonts}
\usepackage{graphicx}
\usepackage{textcomp}
\usepackage{xcolor}
\usepackage{color}
\usepackage{url}
\usepackage{gensymb}
\usepackage[paper=a4paper,top=105pt,bottom=54pt,right=37pt,left=54pt]{geometry}

\def\BibTeX{{\rm B\kern-.05em{\sc i\kern-.025em b}\kern-.08em
    T\kern-.1667em\lower.7ex\hbox{E}\kern-.125emX}}

\begin{document}

\title{\vspace{17pt} Distributed macroscopic traffic \\ simulation with Open Traffic Models}

\author{\IEEEauthorblockN{Gabriel Gomes}
\IEEEauthorblockA{\textit{Institute of Transportation Studies} \\
\textit{University of California}\\
Berkeley, CA, USA \\
gomes@berkeley.edu}
\and
\IEEEauthorblockN{Juliette Ugirumurera}
\IEEEauthorblockA{\textit{Computational Science Center} \\
\textit{National Renewable Energy Laboratory}\\
Denver, CO, USA \\
jugirumu@nrel.gov}
\and
\IEEEauthorblockN{Xiaoye S. Li}
\IEEEauthorblockA{\textit{Computational Research Division} \\
\textit{Lawrence Berkeley National Laboratory}\\
Berkeley, CA, USA \\
xsli@lbl.gov}
}

\maketitle

\begin{abstract}
This paper presents OTM-MPI, an extension of the Open Traffic Models platform (OTM) for running macroscopic traffic simulations in high-performance computing environments. Macroscopic simulations are appropriate for studying regional traffic scenarios when aggregate trends are of interest, rather than individual vehicle traces. They are also appropriate for studying the routing behavior of \textit{classes} of vehicles, such as app-informed vehicles. The network partitioning was performed with METIS. Inter-process communication was done with MPI (message-passing interface). Results are provided for two networks: one realistic network which was obtained from Open Street Maps for Chattanooga, TN, and another larger synthetic grid network. The software recorded a speed-up ratio of 198 using 256 cores for Chattanooga, and 475 with 1,024 cores for the synthetic network. 

\end{abstract}


\IEEEpeerreviewmaketitle

\section{Introduction}
\IEEEPARstart{R}{ecent} years have seen significant technological changes in transportation with the advent of shared-mobility services such as Uber and Lyft, electric vehicles, and vehicle automation. There is a need for government agencies and industry to understand the impact of wide-spread adoption of these new technologies on the transportation system. This in turn requires regional scale modeling tools, which consume large computing and memory resources. Parallel computation and modern supercomputers provide the power and memory to handle such large-scale traffic simulations. They also enable running thousands of parallel simulations that can answer future what-if scenarios.

There are two types of approaches to parallel computation: shared memory and distributed memory.
On a shared-memory computer, parallel simulation takes advantage of the computer's many processors to execute tasks in parallel. In this case, the simulation speed-up is limited by the number of processors on one computer. On a multiprocessor distributed-memory computer system, parallel simulation can use processors on multiple computers while communicating over a message passing system. This paper describes such an approach for a fluid-based traffic simulation model. We begin by reviewing previous efforts, most of which have focused on vehicle-based (microscopic and mesoscopic) models. 

\subsection{Related Works}

\begin{table*}[htbp]
\centering
\begin{tabular} { | p{20mm} | p{20mm} | p{125mm}| } 
	\hline
	\hline
	\textbf{Software} & \textbf{Traffic Model} & \textbf{Parallel Method}\\ \hline
	Paramics \cite{cameron1996paramics}, FastTrans \cite{thulasidasan2009accelerating}    & Microscopic & Parallel computation on distributed-memory computer with MPI for inter-core communication \\ \hline
	TRANSIMS \cite{robertson1969transyt}     & Microscopic & Parallel computation on distributed-memory computer with MPI and PVM for inter-core communication \\ \hline
	AIMSUN \cite{ferrer1993aimsun2}, SEM-Sim \cite{aydt2013multi},  MEgaffic \cite{osogami2012research}     & Microscopic & Multi-threading \\ \hline
	SUMO \cite{behrisch2011sumo}    & Microscopic & Multi-threading for routing \\ \hline
	Dynemo \cite{nokel2002parallel}  & Mesoscopic & Network distribution through master-slave parallelization \\ \hline
	Mobiliti \cite{chan2018mobiliti}     & Mesoscopic & Parallel discrete event with GASNet \\ \hline
	POLARIS \cite{auld2016polaris}     & Mesoscopic & Multi-threading \\ \hline
	BEAM \cite{aboutBeam}    & Mesoscopic & Akka library \\ \hline
	\cite{xu2014mesoscopic,song2017supporting}, \cite{strippgen2009multi}     & Mesoscopic & Parallel computation on GPUs \\ \hline
	\cite{chronopoulos1998real}     & Macroscopic & Distribution of freeway sections on multiple processes of a nCubes2 distributed-memory parallel computer \\ \hline
	\cite{johnston1999parallelization}     & Macroscopic & Distribution of highway segments on multiple processes of a Cray T3W distributed-memory parallel computer \\ \hline
	OTM-MPI \cite{otmmpi}  & Macroscopic & Parallel computing on modern mult-node computer systems with MPI for inter-core communication\\ \hline
\end{tabular}
\caption{Traffic Simulation Software}
\label{tab:softwares}
\end{table*}

Since Greenshield presented the first traffic model in 1934 \cite{greenshields1934photographic}, three main model categories have emerged: microscopic models, macroscopic models, and mesoscopic models. While microscopic models represent the behavior and interaction of individual vehicles,  macroscopic models describe traffic as a continuum. Mesoscopic models are an intermediate class in which macroscopic parameters such as link capacity are applied to individual vehicles. Traffic simulation software tools usually implement a single microscopic, macroscopic or mesoscopic model. Open Traffic Models is one of a few  \textit{hybrid} simulation tools, which allow the combination of two or more models. 

Most of the traffic simulators with parallel computation capabilities are microscopic simulators. An example of this is Paramics \cite{cameron1996paramics}, which was implemented on a the T3D parallel computer with Message Passing Interface (MPI) for inter-processor communication. FastTrans \cite{thulasidasan2009accelerating} and TRANSIMS \cite{nagel2001parallel} are also microscopic simulators and use parallel computation on distributed-memory computer systems by dividing the road network into multiple subnetworks. FastTrans uses MPI to share boundary information among processes, while TRANSIMS can support both MPI and Parallel Virtual Machine (PVM) communication. AIMSUN \cite{ferrer1993aimsun2}, SEMSim \cite{aydt2013multi}, and IBM's Mega MEgaffic simulator \cite{osogami2012research} implement parallel computation with a multi-threading approach, which enables them to update multiple agents simultaneously. The popular microscopic simulator SUMO
exploits multi-threading parallel computation for vehicle routing, but the rest of simulation is single-core \cite{behrisch2011sumo}.
The BEAM simulator \cite{aboutBeam} extends MATSIM \cite{horni2016multi}, which is a microscopic agent-based traffic simulator. BEAM incorporates parallel computation via the Akka \cite{akka} library, which implements the Actor Model paradigm \cite{actorModel}. 

In the realm of mesoscopic traffic simulation, N{\"o}kel and Schmidt present a parallel implementation for Dynemo, that uses a master-slave parallelization. Mobiliti \cite{chan2018mobiliti}, a mesoscopic simulator, models links as agents and vehicles as events. It uses GASNet \cite{gasnet} to enable parallel computation on high performance computing resources. POLARIS  \cite{auld2016polaris} exploits parallel computation via multi-threading. Additional works on mesoscopic traffic simulation with parallel computation on graphics processing units include \cite{xu2014mesoscopic,song2017supporting,strippgen2009multi}. 

On the other hand, parallel computation has not been applied extensively to macroscopic traffic simulation. \cite{chronopoulos1998real} presents a parallel continuum traffic model on an nCubes2 hypercube distributed-memory parallel computer, where a freeway is partitioned into equal segments and assigned to different processors. Similarly, \cite{johnston1999parallelization} describes a parallel lax-momentum traffic model that divided the freeway into equal segments assigned to different processors of a Cray T3E distributed-memory computer. Although  \cite{chronopoulos1998real} and \cite{johnston1999parallelization} use parallel computation for macroscopic traffic simulation on distributed-memory computers, they are focused on freeways only. In addition, the reported studies involve only about 20 miles of freeway, and were executed on older computer systems.

The work presented here is based on the Open Traffic Models simulator. OTM is a full-featured traffic simulator that has in the past been used in single-process environments. OTM \cite{Gomes2019OTM} is a \textit{hybrid} simulator, in the sense that it incorporates multiple (macroscopic, mesoscopic, and microscopic) models which can be combined arbitrarily on a single network. It also supports various road geometries (turn pockets, managed lanes on freeways, etc.), multiple vehicle types, and a variety of traffic control systems (traffic signals, variable speed limits, on-board information systems, etc.). Here we focus on the parallelization of the macroscopic model only. This is both because there are many high quality distributed implementations of vehicle-based models, and because it avoid some of the complexities of asynchronous updates that occur in vehicle-based model. OTM-MPI represents, to the best of our knowledge, the first open-source, distributed-memory, macroscopic simulation model developed for modern high performance parallel machines and large networks  \cite{otmmpi}.

\section{Methodology}
In this section we begin by describing the methodology for dividing the network. We then describe the inter-process communication. Finally we give a brief description of the macroscopic traffic model. 

\begin{figure*}
\centering
  \includegraphics[width=\textwidth]{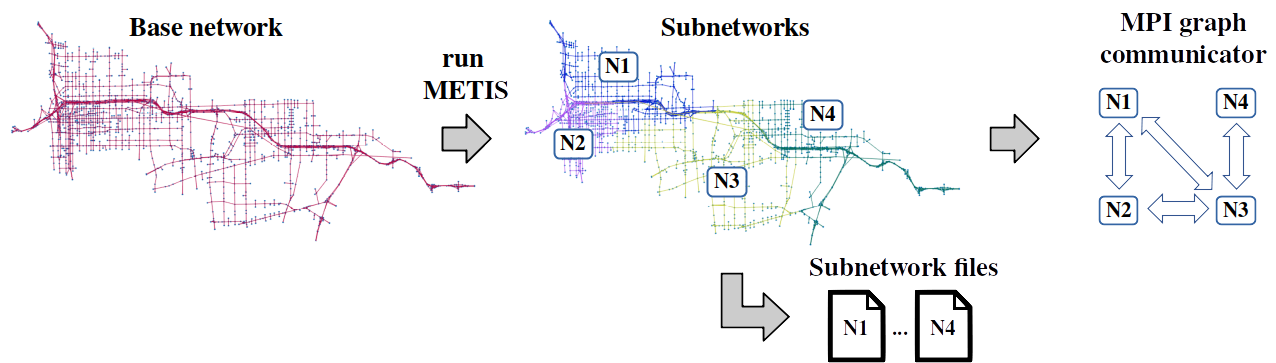}
  \caption{Network partitioning.}
  \label{fig:partition}
\end{figure*}

\subsection{Network partitioning}
The process of distributing the simulation among $n$ cores is illustrated in Figure~\ref{fig:partition}. The first step is to split the base network into $n$ subnetworks. To obtain these, the set of nodes is divided into non-overlapping sets. Let $\mathcal{N}_i$, $i\in[1\hdots n]$, denote the $i$th subset of nodes. The $i$th subnetwork contains all of the nodes in $\mathcal{N}_i$, as well as all of the links with either start or end node in $\mathcal{N}_i$. 
Links with start node in one subset and end node in another constitute the \textit{overlap} between two subnetworks. An overlap link with start node in $\mathcal{N}_i$ and end node in $\mathcal{N}_j$ is called a \textit{relative source} with respect to subnetwork $j$ and a \textit{relative sink} with respect to subnetwork $i$. The road connections that enter (exit) a relative source (sink) link are called \textit{relative source (sink) road connections} (with respect to a given subnetwork). The implementation utilizes an off-line module that invokes METIS~\cite{kaku:98a} to create the node subsets, and then constructs separate input files for each of the subnetworks. These files contain all of the information (and only that information) that is required to run the given subnetwork as a stand-alone simulation. 

\subsection{Process communication}

The connections between the $n$ subnetworks are encoded in a so-called \textit{metagraph}. This is a graph in which each vertex represents the non-overlapping portions of a subnetwork, and the edges are the relative sources and sinks. Each subnetwork is executed by a separate process. The appropriate MPI communicator for this configuration is the \textit{graph communicator}, which encodes the relations of the metagraph. An all-to-all transmission with a graph communicator exchanges messages amongst neighboring vertices in the metagraph. 

The message passed between two neighboring subnetworks consists of an array of numbers representing the vehicles that travel along all of the relative source and sink road connections that join the two subnetworks. During the initialization phase, the processes construct and exchange decoder maps with each of their neighbors. These map each position in the array to a lane group and state index. This same map is used throughout the entire simulation. The approach is conservative in the sense that the size of the message is fixed, and may contain many zeros at any given time. As is shown in Section~\ref{sec:experiments}, the communication time is a small fraction of the total run time, and hence the potential advantage of using dynamic message sizes is small.

The number of vehicles that travel at each time step over relative road connections (i.e. the flow between subnetworks) is computed by the node model in step 2. The MPI communication step is then inserted between steps 2 and 3. This completes the input and output flow computation for relative sources and sinks. After this, step 3) is executed to compute the downstream link for probabilistically routed vehicles, and then the internal state of all of the lane groups is updated. These steps are identical to their counterparts in the sequential model. Hence the result is not affected by the distribution of the calculation over multiple processes.

\subsection{Traffic modeling}
The full mathematical description of OTM can be found in~\cite{Gomes2019OTM}.
The description provided here omits certain details, but is sufficient to understand the inter-process communication that enables distributed simulation. The model is in the class of \textit{cell transmission} models first introduced by Daganzo in \cite{DaganzoCTM}. Since then there have been many extensions and improvements to the CTM, incorporating multiple lanes, non-triangular fundamental diagrams, etc. The CTM model that is implemented in OTM is adapted to an underlying representation of the road network that can be applied to the different  models in the hybrid simulation environment. As with most traffic simulators, there is a high-level graph (links and nodes) which encodes the connectivity of the road network. Each link (i.e. a segment of road between two junctions) has an internal geometry consisting of multiple lanes, turn pockets, and possibly a managed lane with access gates (in the case of a freeway link). There can also be sensors and control devices within the link. 

The relations between lanes that enter and exit a junction are defined by \textit{road connections}. A road connection indicates which lanes in an upstream link can turn into which other lanes in a downstream link. For example, a road connection may indicate that vehicles must be in the outer two lanes of the freeway in order to reach the off-ramp. The road connections that leave a link are used by the program to aggregate lanes into \textit{lane groups}. All lanes in a lane group access the same set of downstream road connections, and are therefore presumed to travel at similar speeds. In our previous example, the two outer lanes of the freeway constitute a lane group, and the remaining inner lanes are another. 

Vehicles in OTM (whether microscopic, mesoscopic, or macroscopic) are classified according to their \textit{type}. The vehicle type encodes two sets of parameters. First, it determines the vehicle's routing behavior which may be deterministic or probabilistic. Deterministically routed vehicles follow a predetermined sequence of links from their origin to their destination. Probabilistically routed vehicles direct themselves at every junction according to time-varying turning probabilities. The software supports arbitrary numbers of deterministic and probabilistic vehicle types. The second parameter of the vehicle type is a \textit{vehicle class}. The vehicle class encodes any set of parameters that the modeler wishes to prescribe. For example, it can be used to distinguish passenger vehicles from trucks (vehicle size), high-occupancy vehicles from single-occupancy vehicles (lane-usage rules), routing app equipped versus not (access to network information), etc.

In OTM's macroscopic model, each lane group is divided into a sequence of \textit{cells}. The state of each cell is the number of vehicles for each vehicle type and each downstream link. The downstream link of a vehicle is the next link in its route. This is determined in step 3) below for probabilistically routed vehicle types. Vehicles change lanes (ie move laterally between cells in adjacent lane groups) in order to reach a lane group that connects (via road connections) to its downstream link. 

The state update for each cell is executed in multiple steps, listed below. The respective formulas can be found in \cite{Gomes2019OTM}.

\vspace{1em} \noindent \textbf{1) Demand and supply calculations.} The total demand and supply for every cell are computed using the standard formulas of the cell-transmission model for a triangular fundamental diagram. For the last cell in each lane group, the demands are computed for each of the road connections that connect to the respective downstream links. Prior to computing the longitudinal demand, the model executes all lateral movements (lane changes).  

\vspace{1em}\noindent \textbf{2) Inter-link flow calculation.} The flows between links travel along road connections. The \textit{node model} resolves these flows from upstream demands and downstream supplies. The details of the node model can be found in \cite{Gomes2019OTM}. Here it will suffice to state that the output of the node model is a set of \textit{vehicle flux packets} that are delivered along road connections to downstream links. The node model ensures that the downstream links have sufficient available space to accommodate the packets.  

\vspace{1em}\noindent \textbf{3) Probabilistic routing.}
When a node has several outgoing links, probabilistically routed vehicles select which one to take according to given turning probabilities. These probabilities are sampled when the vehicle \textit{enters} the approaching link. This is in contrast with many macroscopic models where turning probabilities are sampled at the node, i.e. when the vehicle \textit{exits} the approaching link. This is done in order to improve the representation of lane changing that occurs upstream of the split.

\vspace{1em} \noindent \textbf{4) State update.} Once all packets have been received and tagged with their next downstream link, then the state of the cells can be updated using the conservation of vehicles. This calculation follows the standard cell transmission model formulas for internal cells.

\section{Experiments}
\label{sec:experiments}
We ran OTM-MPI on Cori, a Cray XC40 supercomputer at NERSC~\cite{Cori}. Each of Cray's processing nodes has two sockets, each with a 16-core Intel\textcopyright~Xeon\texttrademark~Processor E5-2698 v3 (``Haswell'') at 2.3 GHz and 128 GB of RAM memory. This is a very modern parallel computer, taking advantage of recent advances in parallel computing technology. We report here two experiments; one with a model of Chattanooga, TN, and another with a large synthetic network. 

\subsection{Experiments with the Chattanooga network}

\begin{figure}[!ht]
\centering
\includegraphics[width=\columnwidth]{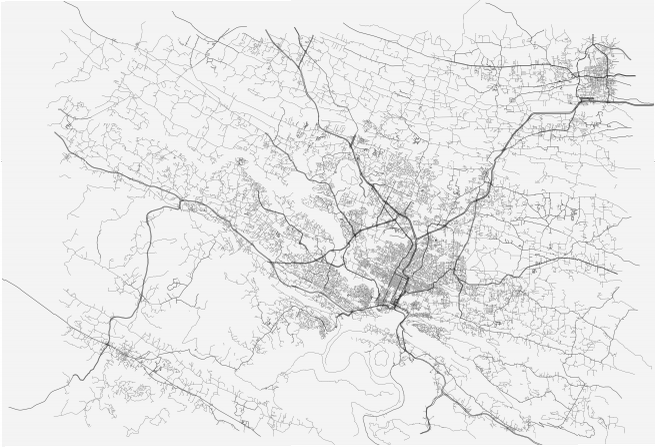}
\caption{OTM model of the Chattanooga, TN. Bounding box: 34.753\degree north, 35.445\degree south, -85.506\degree east, -84.902\degree west. This model has 32,260 nodes and 38,440 links. }
\label{fig:chattanooga}
\end{figure}

\begin{figure}[!ht]
\centering
\includegraphics[width=\columnwidth]{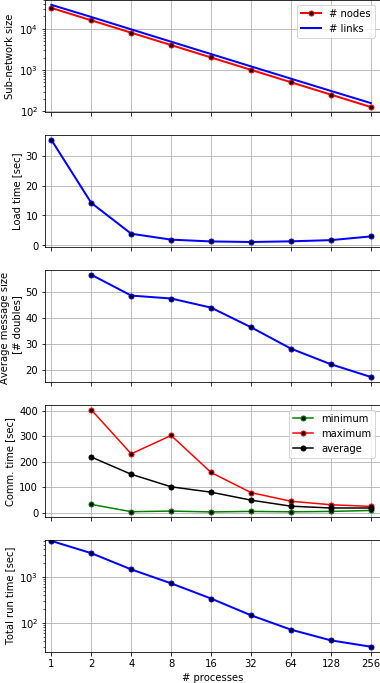}
\caption{Simulation results for the Chattanooga network.}
\label{fig:chattanooga_plots}
\end{figure}

OTM includes a module for extracting networks from Open Street Maps \cite{otmtools,haklay2008openstreetmap}. This module was used to construct a model of  Chattanooga, Tennessee (see Figure~\ref{fig:chattanooga}). The goal of this experiment is to demonstrate the improvements in computation time that can be achieved for a realistic traffic network topology. Traffic performance metrics were not collected for this reason, nor was any effort made to reproduce the demand or routing characteristics of Chattanooga. The network has 32,260 nodes, 38,440 links, and contains 248 source links. Each of these was supplied with a large congestion-inducing demand of 2,000 vph per lane. We conducted 9 runs. In each, the number of parallel processes used was doubled, starting from a single process and up to 256 processes. Each trial was initialized with an empty network and was advanced 10,000 time steps. Figure \ref{fig:chattanooga_plots} shows the results. 

In all of the subplots of the figure, the x-axis is the number of processes involved in the simulation ($n$), represented in base-2 logarithmic scal. The first plot shows that, as expected, the average size of the subnetworks generated by METIS is inversely proportional to the number of processes. The second subplot shows the setup time of the program. This includes the network load time and the time taken to create the MPI communicator and distribute the message decoders. Notice that this load time reaches a minimum and increases slightly for 256 processes. This is because, although the subnetworks are smaller, the complexity of the metagraph and decoders increases and becomes significant for $n=256$. 

The third plot shows the average size of the messages being passed through the MPI communicator. For $n=2$, each process sends an average of 56 floats to each of its neighbors in the metagraph. As $n$ increases, the size of the messages gets smaller, while the number of messages and the total amount of information increases. The fourth subplot shows that the total time spent in communication generally \textit{decreases} with larger $n$. The graph communicator can more quickly transmit many small messages than a few large messages. This trend is true on average, however the plot also shows that the maximum and minimum communication times are not monotonic. 

Finally the fifth subplot shows the total run time, including load, communication, and computation time, on a log-log scale. The nearly linear trend indicates an inverse proportional relation between run time and the number of processes.  For this network, the execution on a single process took 6,026 seconds (1 hour and 40 minutes), and was reduced to 30.6 seconds when run with 256 cores. This corresponds to speed up ratio of 198.
 
\subsection{Experiment with a synthetic network}
\begin{figure}[!ht]
    \centering
    \includegraphics[height=0.5\columnwidth]{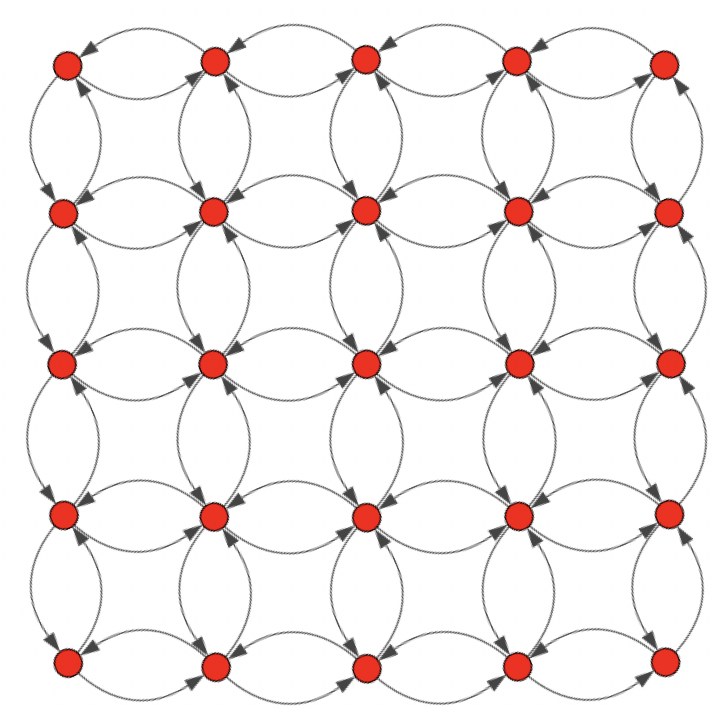}
    \caption{Topology of a synthetic network with 81,250 nodes and 268,000 links.}
    \label{fig:Synthetic_Network}
\end{figure}

We conducted experiments on a large, grid-like, synthetic network, with a tiled topology as shown in Figure~\ref{fig:Synthetic_Network}. The goal of this experiment was to test the program on a second larger network, this one with 81,250 nodes and 268,000 links, which is about seven times the size of the Chattanooga network (in terms of links). 

\begin{figure}[!ht]
    \centering
    \includegraphics[width=\columnwidth]{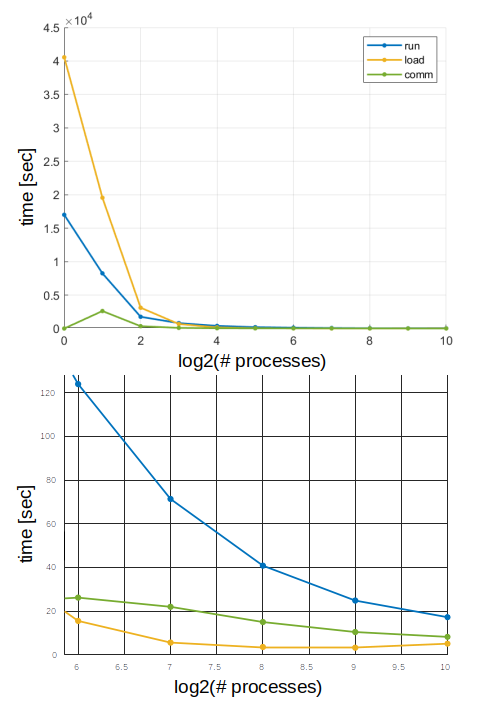}
    \caption{Results with the grid-like network.}
    \label{fig:mpirun}
\end{figure}

Source flows were placed on 15,620 source links, again in sufficient amount to  create congestion. The simulation was run with $n$ ranging from 1 to 1024 processes (11 runs), and run for 1,000 simulated seconds each. Figure \ref{fig:mpirun} shows load time, communication time, and computation (run) time in each case. 
The lower plot focuses on the simulations that are not distinguishable in the upper plot ($n\geq 6$). The trend is similar as in the first example, with inverse proportionality between the run time and the number of processes. 

\begin{figure}[!ht]
    \centering
    \includegraphics[width=\columnwidth]{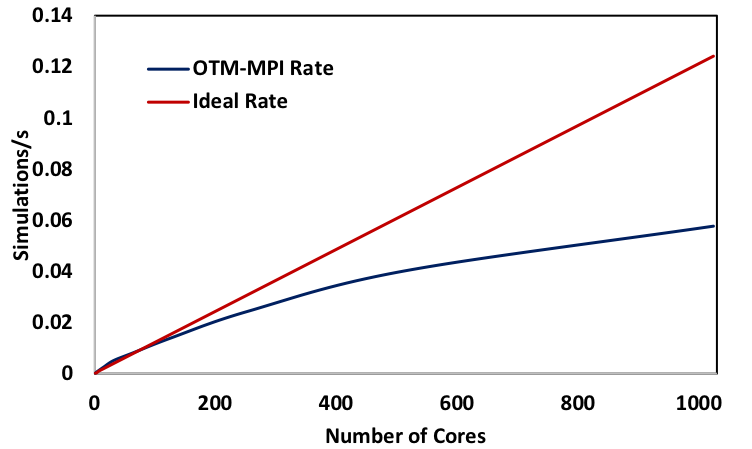}
    \caption{OTM scaling  compared to ideal scaling for the grid network.}
    \label{fig:scaling}
\end{figure}

Figure~\ref{fig:scaling} demonstrate the scaling of OTM-MPI compared to the ideal scaling simulation rate. For each simulation experiment, the simulation rate $1/(simulation \:time)$, which corresponds to the number of simulations that can be completed within 1 seconds (simulations/s). The ideal rate is calculated as $(number\: of\: processes)*(serial\: simulation\: rate)$, and assumes that simulation rate increases proportionally with the number of processes. In practice, parallel simulation does not attain the ideal scaling rate as the $(communication\:time)/(simulation\:time)$ ratio grows with the number of processes. In fact, we observed that for the experiment with 1024 processes, each process spend half of its computation time in communicating with other processes. However, we still observed that for the simulation time, the parallel OTM had a speed up of 475 times with 1024 processes compared to the serial OTM. This corresponded to a time reduction from 8,245 seconds to 17 seconds.

\section{Conclusion}
The main contribution of this work is an open-source implementation of macroscopic traffic simulation for high-performance computing (HPC). As was detailed in the introduction, there are currently a host of platforms for running vehicle-based simulations in HPC. OTM-MPI offers an alternative for scenarios in which macroscopic modeling may be preferred over vehicle-based simulation. This may be the case if the following two conditions are given. First, that the available data is better suited for estimating macroscopic quantities (capacities) than microscopic ones (car-following rules). For example, if the data consists in loop detector measurements, and detailed vehicle accelerations are absent. Second, that the metrics of interest can be obtained from the macroscopic model. For instance, if we are interested in travel times and delays, and not in vehicle accelerations. Studies of routing behaviors in particular, either `natural' or induced by routing apps, can equally be carried out in macroscopic models. OTM-MPI can be obtained from \cite{otmmpi}.

\section*{Acknowledgement}
This work was supported in part by the Office of Science of
the U.S. Department of Energy under contract No. DE-AC02-
05CH11231. This work was authored in part by the National Renewable Energy Laboratory, operated by Alliance for Sustainable Energy, LLC, for the U.S. Department of Energy under Contract No. DE-AC36-08GO28308. 


\bibliography{_refs}{}

\begin{thebibliography}{10}
\providecommand{\url}[1]{#1}
\csname url@samestyle\endcsname
\providecommand{\newblock}{\relax}
\providecommand{\bibinfo}[2]{#2}
\providecommand{\BIBentrySTDinterwordspacing}{\spaceskip=0pt\relax}
\providecommand{\BIBentryALTinterwordstretchfactor}{4}
\providecommand{\BIBentryALTinterwordspacing}{\spaceskip=\fontdimen2\font plus
\BIBentryALTinterwordstretchfactor\fontdimen3\font minus
  \fontdimen4\font\relax}
\providecommand{\BIBforeignlanguage}[2]{{%
\expandafter\ifx\csname l@#1\endcsname\relax
\typeout{** WARNING: IEEEtran.bst: No hyphenation pattern has been}%
\typeout{** loaded for the language `#1'. Using the pattern for}%
\typeout{** the default language instead.}%
\else
\language=\csname l@#1\endcsname
\fi
#2}}
\providecommand{\BIBdecl}{\relax}
\BIBdecl

\bibitem{cameron1996paramics}
G.~D. Cameron and G.~I. Duncan, ``{PARAMICS, Parallel microscopic simulation of
  road traffic},'' \emph{The Journal of Supercomputing}, vol.~10, no.~1, pp.
  25--53, 1996.

\bibitem{thulasidasan2009accelerating}
S.~Thulasidasan and S.~Eidenbenz, ``Accelerating traffic microsimulations: A
  parallel discrete-event queue-based approach for speed and scale,'' in
  \emph{Winter Simulation Conference}.\hskip 1em plus 0.5em minus 0.4em\relax
  Winter Simulation Conference, 2009, pp. 2457--2466.

\bibitem{robertson1969transyt}
D.~I. Robertson, ``{TRANSYT: a traffic network study tool},'' \emph{RRL Report
  LR 253, Road Research Laboratory.}, 1969.

\bibitem{ferrer1993aimsun2}
J.~L. Ferrer and J.~Barcel{\'o}, ``{AIMSUN2: Advanced interactive microscopic
  simulator for urban and non-urban networks},'' \emph{Research Report}, 1993.

\bibitem{aydt2013multi}
H.~Aydt, Y.~Xu, M.~Lees, and A.~Knoll, ``{A multi-threaded execution model for
  the agent-based SEMSim traffic simulation},'' in \emph{Asian Simulation
  Conference}.\hskip 1em plus 0.5em minus 0.4em\relax Springer, 2013, pp.
  1--12.

\bibitem{osogami2012research}
T.~Osogami, T.~Imamichi, H.~Mizuta, T.~Morimura, R.~Raymond, T.~Suzumura,
  R.~Takahashi, and T.~Id, ``{Research Report IBM Mega Traffic Simulator},''
  Technical report, Tech. Rep., 2012.

\bibitem{behrisch2011sumo}
M.~Behrisch, L.~Bieker, J.~Erdmann, and D.~Krajzewicz, ``{SUMO - Simulation of
  urban mobility},'' in \emph{The Third International Conference on Advances in
  System Simulation (SIMUL 2011), Barcelona, Spain}, vol.~42, 2011.

\bibitem{nokel2002parallel}
K.~N{\"o}kel and M.~Schmidt, ``{Parallel DYNEMO: Meso-scopic traffic flow
  simulation on large networks},'' \emph{Networks and Spatial Economics},
  vol.~2, no.~4, pp. 387--403, 2002.

\bibitem{chan2018mobiliti}
C.~Chan, B.~Wang, J.~Bachan, and J.~Macfarlane, ``Mobiliti: Scalable
  transportation simulation using high-performance parallel computing,'' in
  \emph{To appear: IEEE International Conference on Intelligent Transportation
  Systems (ITSC)(November 2018)}, 2018.

\bibitem{auld2016polaris}
J.~Auld, M.~Hope, H.~Ley, V.~Sokolov, B.~Xu, and K.~Zhang, ``{POLARIS:
  Agent-based modeling framework development and implementation for integrated
  travel demand and network and operations simulations},'' \emph{Transportation
  Research Part C: Emerging Technologies}, vol.~64, pp. 101--116, 2016.

\bibitem{aboutBeam}
C.~Sheppard and R.~Waraich, ``{About BEAM},''
  \url{https://github.com/LBNL-UCB-STI/beam}, accessed: 2020-03-01.

\bibitem{xu2014mesoscopic}
Y.~Xu, G.~Tan, X.~Li, and X.~Song, ``{Mesoscopic traffic simulation on
  CPU/GPU},'' in \emph{Proceedings of the 2nd ACM SIGSIM Conference on
  Principles of Advanced Discrete Simulation}.\hskip 1em plus 0.5em minus
  0.4em\relax ACM, 2014, pp. 39--50.

\bibitem{song2017supporting}
X.~Song, Z.~Xie, Y.~Xu, G.~Tan, W.~Tang, J.~Bi, and X.~Li, ``{Supporting
  real-world network-oriented mesoscopic traffic simulation on GPU},''
  \emph{Simulation Modelling Practice and Theory}, vol.~74, pp. 46--63, 2017.

\bibitem{strippgen2009multi}
D.~Strippgen and K.~Nagel, ``{Multi-agent traffic simulation with CUDA},'' in
  \emph{High Performance Computing \& Simulation, 2009. HPCS'09. International
  Conference on}.\hskip 1em plus 0.5em minus 0.4em\relax IEEE, 2009, pp.
  106--114.

\bibitem{chronopoulos1998real}
A.~T. Chronopoulos and C.~M. Johnston, ``A real-time traffic simulation
  system,'' \emph{IEEE Transactions on Vehicular Technology}, vol.~47, no.~1,
  pp. 321--331, 1998.

\bibitem{johnston1999parallelization}
C.~M. Johnston and A.~T. Chronopoulos, ``The parallelization of a highway
  traffic flow simulation,'' in \emph{Frontiers of Massively Parallel
  Computation, 1999. Frontiers' 99. The Seventh Symposium on the}.\hskip 1em
  plus 0.5em minus 0.4em\relax IEEE, 1999, pp. 192--199.

\bibitem{otmmpi}
``{OTM-MPI},'' \url{https://github.com/ggomes/otm-mpi}, accessed: 2020-03-01.

\bibitem{greenshields1934photographic}
B.~D. Greenshields, J.~Thompson, H.~Dickinson, and R.~Swinton, ``The
  photographic method of studying traffic behavior,'' in \emph{Highway Research
  Board Proceedings}, vol.~13, 1934.

\bibitem{nagel2001parallel}
K.~Nagel and M.~Rickert, ``{Parallel implementation of the TRANSIMS
  micro-simulation},'' \emph{Parallel Computing}, vol.~27, no.~12, pp.
  1611--1639, 2001.

\bibitem{horni2016multi}
A.~Horni, K.~Nagel, and K.~W. Axhausen, \emph{The multi-agent transport
  simulation MATSim}.\hskip 1em plus 0.5em minus 0.4em\relax Ubiquity Press
  London:, 2016.

\bibitem{akka}
``akka,'' \url{https://akka.io/}, accessed: 2020-03-01.

\bibitem{actorModel}
``Actor model,''
  \url{https://en.wikipedia.org/wiki/Actor_model#Message-passing_semantics},
  accessed: 2020-03-01.

\bibitem{gasnet}
``{GASNet},'' \url{https://gasnet.lbl.gov/}, accessed: 2020-03-01.

\bibitem{Gomes2019OTM}
G.~Gomes, ``{Open Traffic Models - A framework for hybrid simulation of
  transportation networks},'' \emph{Submitted for consideration to the Journal
  of Simulation.}, 2019.

\bibitem{kaku:98a}
G.~Karypis and V.~Kumar, ``{METIS - Serial graph partitioning and computing
  fill-reducing matrix ordering},'' university of Minnesota.
  \url{http://glaros.dtc.umn.edu/gkhome/views/metis}.

\bibitem{DaganzoCTM}
C.~Daganzo, ``The cell transmission model: A dynamic representation of highway
  traffic consistent with the hydrodynamic theory,'' \emph{Transportation
  Research Part B: Methodological}, vol.~28, no.~4, pp. 269--287, 1994.

\bibitem{Cori}
``Cori,'' \url{http://www.nersc.gov/users/computational-systems/cori/},
  accessed: 2020-03-01.

\bibitem{otmtools}
``{OTM Tools},'' \url{github.com/ggomes/otm-tools}, accessed: 2020-03-01.

\bibitem{haklay2008openstreetmap}
M.~Haklay and P.~Weber, ``{OpenStreetMap: User-generated street maps},''
  \emph{IEEE Pervasive Computing}, vol.~7, no.~4, pp. 12--18, 2008.

\end{thebibliography}
\bibliographystyle{IEEEtran}

\end{document}